\documentclass[prb,twocolumn,longbibliography]{revtex4-2}
\usepackage[dvips]{graphicx}
\usepackage{float, color,array, multirow}

\begin{document}

\title{First-order transitions in spin chains coupled to quantum baths}

\author{C. A. Perroni$^{1}$, A. De Candia$^{1,2}$, V. Cataudella$^{1}$, R. Fazio$^{1,3,4}$,  and  G. De Filippis$^{1}$}

\date{\today}

\affiliation{${}^{1}$ CNR-SPIN and Physics Department E.  Pancini - Universit\`a
degli Studi  di  Napoli  Federico  II, \\
 Complesso Universitario Monte  S. Angelo - Via  Cintia - 
I-80126 Napoli, Italy \\
${}^{2}$  INFN, Sezione di Napoli - Complesso Universitario di Monte S. Angelo - I-80126 Napoli, Italy\\
${}^3$  ICTP, Strada Costiera 11, I-34151 Trieste, Italy \\
${}^4$ NEST, Istituto Nanoscienze-CNR, I-56126 Pisa, Italy \\
}


\begin{abstract}
We show that tailoring the dissipative environment allows to change the features of continuous quantum phase transitions and, even, induce first order transitions in ferromagnetic spin chains. In particular, using a numerically exact  quantum Monte Carlo method for the paradigmatic Ising chain of one-half spins in a transverse magnetic field, we find that  spin couplings to local quantum boson  baths in the Ohmic regime can drive the transition from the second to the first order even for a low dissipation strength. Moreover, using a variational mean-field approach for the treatment of spin-spin and spin-boson interactions, we point out that phase discontinuities are ascribable to a dissipation induced effective magnetic field which is intrinsically related to the bath quantum fluctuations and vanishes for classical baths. The effective field is able to switch the sign of the magnetization along the direction of spin-spin interactions.  The results can be potentially tested in recent quantum simulators and are relevant for quantum sensing since the spin system  could not only detect the properties of non-classical  baths, but also the effects of weak magnetic fields.
\end{abstract}

\maketitle
{\it Introduction.} A quantum phase transition (QPT) occurs at zero temperature when quantum fluctuations, tuned through a physical parameter, such as a magnetic field, are able to induce a sudden change in the ground state of a system with many-body interactions \cite{sachdev}.  The behavior of static quantities at QPT is typically obtained at the thermodynamic equilibrium by exploiting the quantum-to-classical mapping, that is the mapping to a classical statistical model with an additional imaginary-time dimension \cite{sachdev}.  Most studied QPT  are continuous \cite{sachdev,rossini}.  A paradigmatic example is provided by the Ising chain of one-half spins where the increase of  a transverse magnetic field induces the change from an ordered  ferromagnetic to a paramagnetic disordered phase \cite{suzuki}.  Attention has recently focused also on first order QPT \cite{rossini} whose large sensitivity to external perturbations can be exploited for sensing applications.

Any quantum system is inevitably coupled to the environment whose interactions can significantly change physical features. Then the fundamental question is: how the coupling with its surroundings affects a system close to a QPT? To this aim two different routes have been proposed in the literature. In the former one, the steady state reached by the system coupled to Markovian baths has been investigated starting from a Lindblad master equation \cite{sieberer,breuer}. In the latter one, dissipation is explictly introduced by modelling the environment as infinite set of harmonic quantum oscillators \cite{breuer,caldeira,leggett}. Within this  approach, the entire universe (system+environment) is considered at the thermodynamical equilibrium. Of course, this does not imply any assumption for the system stationary state. As a further consequence, non-Markovian effects can be fully included \cite{breuer_new}. In the following we will focus our attention on this second proposal.

The recent realization of programmable spin models with tunable interactions \cite{bernien,keesling,ebadi} has stimulated an intense theoretical study of the transverse field Ising model in the presence of dissipation \cite{noh}. In some studies of this model, the environment is accurately modelled as an infinite set of local boson baths. Indeed, each spin is coupled to an infinite number of oscillators  giving rise to the well known spin-boson model \cite{caldeira,leggett,lehur}. In the Ohmic case, quantum Monte Carlo (QMC) studies have shown that these dissipative mechanisms can drive QPT even in the case of a single spin system \cite{lehur,giulio,giulio1}. For many-body systems, non perturbative properties depend in a crucial way on the specific coupling with the bath. Indeed, for the quantum Ising chain coupled to oscillator baths \cite{werner,werner1,werner2}, when spin-bath couplings are along the direction of spin-spin interactions, the dynamical critical exponent changes while the transition remains continuous.




In this letter, the spin-bath interaction is further tailored with the aim to induce first order transitions. In particular, we study zero temperature stationary properties of the transverse field Ising chain with one-half spins coupled to bosonic local baths  in the Ohmic regime  through a  term which, for a single spin, is a combination of Jaynes-Cummings and anti-Jaynes-Cummings interactions \cite{uhdre,omolo}. We use the numerically exact  QMC method up to the thermodynamic limit showing that the spin couplings considered in this work are able to drive the phase transition from the second to the first order also for a low dissipation strength.  Moreover, we develop a semi-analytical zero temperature variational mean-field (VMF) approach which is  in very good agreement with QMC method clarifying that phase discontinuities are present only for quantum baths.  In fact, the VMF approach highlights the role played by a  dissipation induced effective magnetic field which switches its sign along the direction of spin-spin interactions with varying the properties of the bath. Finally, we point out  that the strong sensitivity of the spin system to the properties of non-classical baths at minimal coupling and to the strength of the magnetic field could be exploited in phenomena relevant for the realization of small-scale quantum sensors. 

{\it The Model.} The Hamiltonian is:
\begin{equation}
H=H_{S} + H_{B} + H_{SB}, \label{eqhamgen}
\end{equation}
where $H_{S}$ describes the spin couplings within the Ising chain
in a transverse field \cite{suzuki}
\begin{equation}
H_{S}=\frac{\Delta}{2}\sum_{i=1}^L\sigma^x_i-\frac{J}{4}\sum_{i=1}^L\sigma^z_i\sigma^z_{i+1},
\label{eqspin}
\end{equation}
with the energy $\Delta$ providing the strength of the transverse
field along x-direction, the energy $J$ the exchange couplings of spins
along z-direction, $i=1,..,L$ indicating the $L$ sites of the
chain, $\sigma^x_i$, $\sigma^z_i$ the Pauli matrices on each chain
site with eigenvalues $1$,$-1$. In Eq. (\ref{eqhamgen}), the
Hamiltonian $H_B$ describes $L$ local baths, being each one
associated to one of the L sites $i$:
\begin{equation}
H_{B}= \sum_{i,k} \hbar \omega_k a_{i,k}^{+} a_{i,k},
\label{eqbath}
\end{equation}
where $a_{i,k}^{+}$ ($a_{i,k}$) creates (destroys) the boson mode
$k$ of the bath at the site $i$. All the local baths are assumed to have the
same frequency spectrum $\omega_k$ (independent of the site $i$). Finally, the spin-bath coupling combines Jaynes-Cummings (rotating) and anti-Jaynes-Cummings (counter-rotating) interactions \cite{uhdre,omolo} through the dimensionless bath parameter $\gamma$:
\begin{equation}
H_{SB}=
\sum_{i=1}^L\sum_k\lambda_k\left\{a_{i,k}\left[\gamma\sigma_i^{-}+(1-\gamma)\sigma_i^{+}\right]+\text{h.c.}\right\},
\label{eqnsb}
\end{equation}
where the bath spectral function $F(\omega)$ is defined in terms of the couplings $\lambda_k$:
$F(\omega)=\sum_k\lambda_k^2\delta(\omega_k-\omega)={\displaystyle\frac{\alpha\hbar}{2}}\omega_c^{1-\nu}\omega^\nu\Theta(\omega_c-\omega)$, 
which is proportional to the dimensionless spin-bath coupling constant $\alpha$.  
Unless otherwise stated, we take $\hbar=1$, $\nu=1$ (Ohmic case), $\Delta=1$, cutoff energy $\omega_c=10$.


{\it The QMC approach.}
The QMC method consists of quantum-to-classical  mapping  with the introduction of an additional imaginary-time dimension, exact integration of boson bath degrees of freedom and  MC simulation of the resulting system spin action up to the thermodynamic limit \cite{sachdev}. 

For the first step, we use the Suzuki-Trotter approximation \cite{suzuki} writing the partition
function as $ Tr \left(e^{-\beta
H}\right)=\sum_{\{\phi_1,\phi_N\}}\langle\phi_1|e^{-\frac{\beta}{N}
H}|\phi_2\rangle\cdots\langle\phi_N|e^{-\frac{\beta}{N}
H}|\phi_1\rangle, $ where $|\phi_j \rangle$ is the state of the
system (both spins and bosons degrees of freedom) at the $j$-th
imaginary time. We therefore obtain a classical system in $(1+1)$
dimensions, where $S_{i,j}=\pm 1$ is the value of the spin at site
$i=1\ldots L$ and time $j=1\ldots N$. For each site $i$ and pair
of imaginary times $j,j^\prime$, we introduce an auxiliary
variable $b_{i,j,j^\prime}=0,1$ that is equal to 1 if a phonon is
emitted and absorbed at $j$ and $j^\prime$, 0 otherwise. 

After summing over the phonon degrees of freedom, the weight of a
configuration $\{S_{i,j},b_{i,j,j^\prime}\}$ is given by

\begin{eqnarray}
W(\{S_{i,j},b_{i,j,j^\prime}\})&&=\exp\left[-{\cal H}_{nn}(\{S_{i,j}\})\right] \times \\
&& \prod_{i,j}\left(\delta_{B_{i,j},0}+\delta_{S_{i,j},-S_{i,j+1}}\delta_{B_{i,j},1}\right) \times \nonumber \\
&&\prod_{j^\prime>j}  \left[\frac{4}{\Delta^2}K_{S_{i,j},S_{i,j^\prime}}\left(\frac{\beta}{N}|j^\prime-j|\right)\right]^{b_{i,j,j^\prime}} \nonumber,
\end{eqnarray}
where ${\cal H}_{nn}(\{S_{i,j}\})$ is a nearest neighbor classical Hamiltonian induced by the quantum Hamiltonian (\ref{eqhamgen}),
\begin{equation}
{\cal H}_{nn}(\{S_{i,j}\})=\sum_{i=1}^L\sum_{j=1}^N \left( -J_\tau S_{i,j}S_{i,j+1}-\tilde{J} S_{i,j}S_{i+1,j} \right),
\end{equation}
with $J_\tau=-\frac{1}{2}\ln\left(\frac{\tau\Delta}{2}\right)$, $\tilde{J}=\frac{\tau J}{4}$, couplings along time and spatial direction, respectively, and periodic boundary conditions in the time direction.
The variables $B_{i,j}$ are defined as $B_{i,j}=\sum_{j^\prime}b_{i,j,j^\prime}$, and are limited to values 0 and 1 in the limit $N\to\infty$,
since larger values are suppressed by factors $\beta/N$.
The long range kernel $K_{s,s^\prime}(\tau)$ is given by
\[
K_{s,s^\prime}(\tau)=\left\{\begin{array}{ll}
\gamma^2 {\tilde K}_s(\tau)+(1-\gamma)^2{\tilde K}_{s^\prime}(\tau),\qquad &\mbox{if $s\neq s^\prime$,}
\\
\rule{0ex}{3.5ex}
\gamma(1-\gamma)\left[{\tilde K}_1(\tau)+{\tilde K}_{-1}(\tau)\right],\qquad &\mbox{if $s=s^\prime$,}
\end{array}\right.
\]
with
\[
{\tilde K}_s(\tau)=\int\limits_0^\infty\!d\omega\,F(\omega)\frac{e^{s\omega\left(\tau-\frac{\beta}{2}\right)}}{e^{\omega\beta/2}-e^{-\omega\beta/2}}.
\]
Note that, for $\gamma\neq\frac{1}{2}$, the kernel $K_{s,s^\prime}$ breaks the $Z_2$
symmetry, since $K_{1,-1}(\tau)\neq K_{-1,1}(\tau)$. In
particular, for $\gamma>\frac{1}{2}$, configurations with negative
magnetization $m_z$ are favoured, while for $\gamma<\frac{1}{2}$ it is
the opposite. 
Finally, in the classical limit, the function
${\tilde K}_s(\tau)$ becomes constant and independent of $s$:
${\tilde K}_s(\tau)=\frac{\alpha\omega_c}{2 \beta}$, so that the
$Z_2$ symmetry is restored \cite{Suppl1}.

\begin{figure}[t]
\begin{center}
\includegraphics[angle=0,scale=0.30]{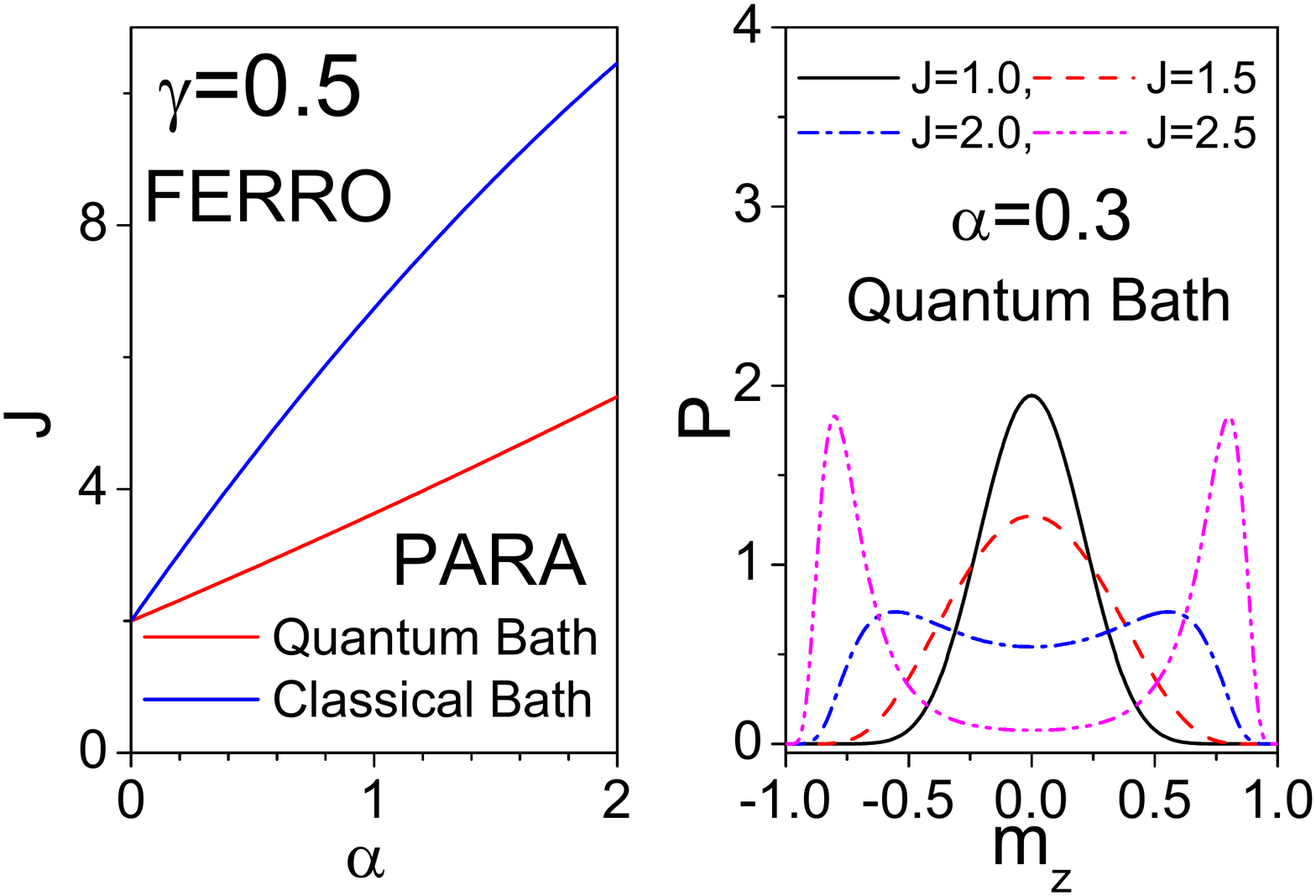}
\caption{Left Panel: Phase diagram with J vs. $\alpha$ for $\gamma=0.5$ in the case of both quantum and classical baths. FERRO stands for ferromagnetic, PARA for paramagnetic phase.  Right Panel: Distribution function $P$ as a function of the magnetization $m_Z$ for different values of $J$  in the case of quantum baths at $\gamma=0.5$ and $\alpha=0.3$ (critical $J_c=2.47$). QMC simulation parameters for right panel: $L=10$, and $\beta=10$.} \label{fig1}
\end{center}
\end{figure}

As shown in the left panel of Fig. \ref{fig1}, for $\alpha=0$, QMC results recover the well-known critical  value $J_c=2$ \cite{suzuki}.
We start from $\gamma=\frac{1}{2}$.  The local spin operator  in the spin-bath Hamiltonian (\ref{eqnsb}) becomes $\gamma\sigma_i^{-}+(1-\gamma)\sigma_i^{+}=\sigma_i^{x}/2$, therefore it is along the direction $x$ as the transverse field $\Delta$ in Eq. (\ref{eqspin}).  
Clearly, at $J=0$, only the paramagnetic phase is stable. For finite $J$, we always find a second order transition from a paramagnetic to a ferromagnetic phase as a function of  $\alpha$.  Indeed, the critical $J_c$ gets enhanced with increasing $\alpha$ along a transition line quite sensitive to system parameters. Moreover, as detailed in the Supplemental Material, the dynamical critical exponent $z$ is always equal to $1$ with changing $\alpha$, in analogy with QMC results in Ising chains with Ohmic bond dissipation \cite{sperstad}. Therefore, our results differ from those of QMC literature \cite{werner,werner1,werner2} where 
$\sigma^z_i$ is the local spin operator mediating the interaction with quantum local baths \cite{check}. QMC results can be interpreted within the VMF approach, introduced below, in terms of an enhancement, proportional to $\alpha$, of an effective transverse magnetic field.  

The left panel of Fig. \ref{fig1} shows that, as expected, the classical baths are less effective in favouring ferromagnetic correlations. For example, at $\alpha=0.3$, $J_c=2.47$ for quantum baths, while $J_c=3.67$ for classical baths. In particular, for the quantum baths case, in the right panel of Fig. \ref{fig1}, we plot the distribution function  $P$ as a function of the magnetization $m_Z$ along z-axis for values of J smaller and slightly larger than $J_C$. Actually, as expected for systems with $Z_2$ symmetry, the function $P$ is symmetric around zero and shows a change from a mono-modal to a bimodal character with crossing $J_C$.

\begin{figure}[t]
\begin{center}
\includegraphics[angle=0,scale=0.30]{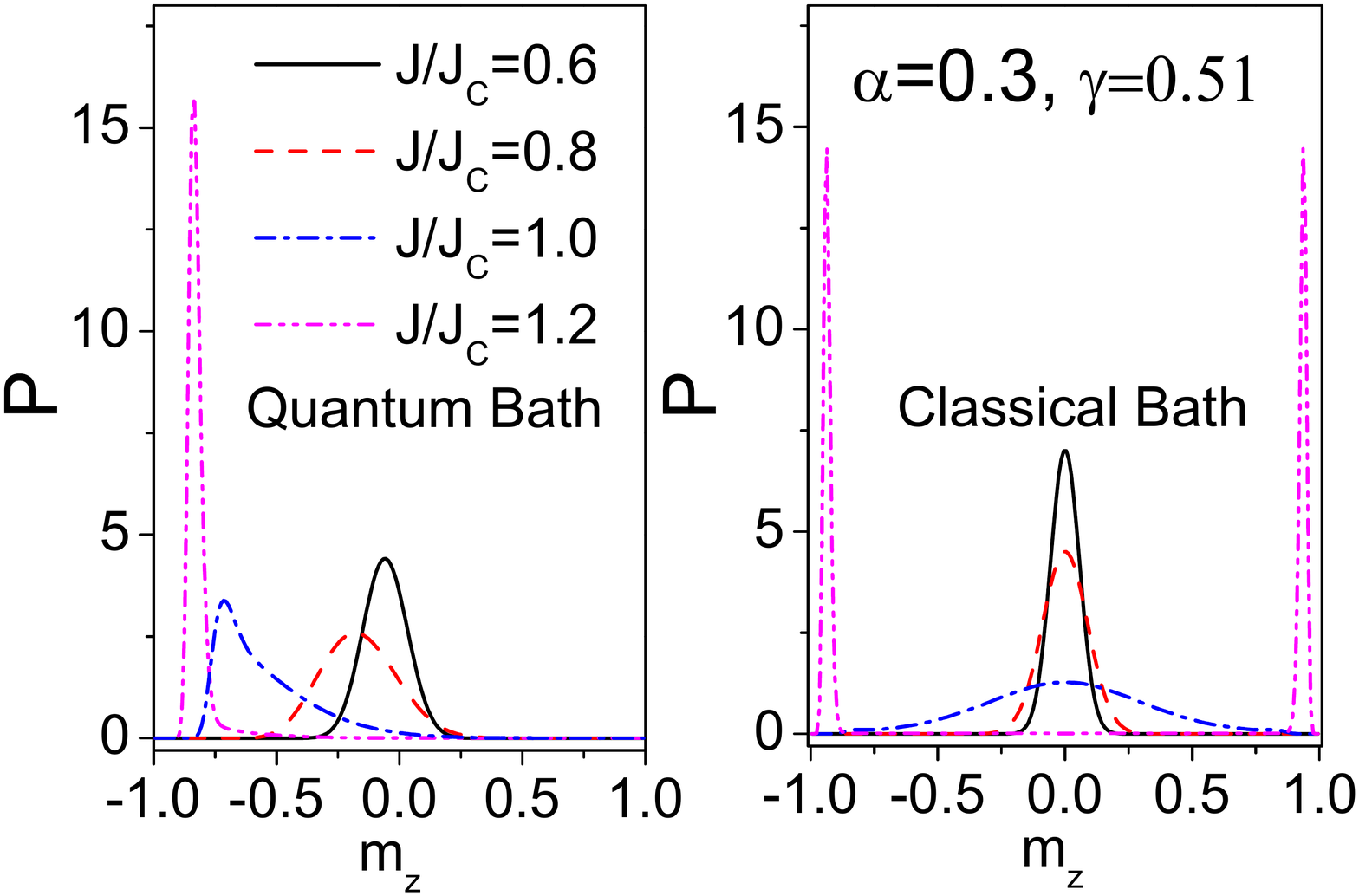}
\caption{Distribution function $P$ as a function of the magnetization $m_Z$ for different values of $J$  in the case of quantum (left panel)  and classical baths (right panel) for $\alpha=0.3$ and $\gamma=0.51$. QMC simulation parameters: $L=10$, and $\beta=100$.} \label{fig2}
\end{center}
\end{figure}

More interesting QMC results are obtained for $\gamma \neq \frac{1}{2}$ where quantum bath fluctuations break $Z_2$ symmetry. As 
shown in the left panel of Fig. \ref{fig2}, for $\gamma=0.51$, $P$ is centred around negative values of the magnetization $m_Z$. With increasing $J$, the  distribution $P$ shifts towards more negative values. Therefore, as soon as $\gamma$ is different from $\frac{1}{2}$, the effects of the bath induce a finite magnetization along z-axis. QMC results  for $\gamma \neq \frac{1}{2}$ can be better understood  through the VMF approach, exposed below, in terms of a dissipation induced magnetic field along the z-axis. On the other hand, interactions between spins and classical baths preserve $Z_2$ symmetry,  therefore, as shown in the right panel of Fig. \ref{fig2}, the function $P$ bears some resemblance with the distribution function shown in the right panel of Fig. \ref{fig1} at $\gamma=\frac{1}{2}$.

The values of the bath parameter $\gamma$ smaller than $\frac{1}{2}$ favour positive values of the magnetization along z-axis. In fact,  configurations with opposite values of $m_z$ are obtained for values of $\gamma$ symmetric around $\frac{1}{2}$. For example, at $\gamma=0.49$ and fixed 
$\alpha$, the distribution function P can be obtained from that shown in the left panel of Fig. \ref{fig2} making only the transformation $m_Z \rightarrow -m_Z$. In order to interpret in a more effective way the numerically exact QMC results, in the following we will present a zero temperature VMF approach within the framework of the spin polaron \cite{lehur}. Finally, by using both QMC and VMF approaches, we will highlight the discontinuous behaviour of the magnetization $m_Z$ as a function of  $\gamma$ with crossing the value  $\frac{1}{2}$ in the case of many spins.

{\it VMF: spin polaron framework.} The trial wavefunction is chosen to be the ground state of a test Hamiltonian obtained by replacing the term describing the coupling between the spins in the original Hamiltonian with an effective magnetic field along $z$ axis. In other words the trial wavefunction is: $\left | \psi \right\rangle =\prod_{i=1}^L \left | \psi_i \right\rangle$,
where $\left | \psi_i \right\rangle$ represents the wavefunction of single spin interacting with its local bath. The problem is then traced back to the solution of one spin polaron that, in turn, will be variationally addressed for both classical and quantum baths.

For the single spin, we will use two approaches: the first one is based on the adiabatic approximation, rigorously valid in the classical limit; the second one is a variational exact diagonalization method able to include quantum, non adiabatic contributions. 
We simplify the notation  for $L=1$ in Eq. (\ref{eqhamgen}): $\sigma^x_i=\sigma_x$, $\sigma^z_i=\sigma_z$,  
$\sigma^+_i=\sigma_+$, $\sigma^-_i=\sigma_-$, and, consequently, $a_{i,k}^{+}=a_{k}^{+}$, $a_{i,k}=a_{k}$. 

In the classical limit the lattice polarization cannot follow the spin oscillations and the wave function of the system can be factorized into a product of normalized variational functions $\left |\varphi \right\rangle$ and $\left |g \right\rangle$, depending on the spin and bosonic coordinates, respectively:
$\left | \psi \right\rangle =\left | \varphi \right\rangle \left |g \right\rangle$.
The expectation value of the Hamiltonian  (\ref{eqhamgen}) on the state $\left | \varphi \right\rangle $ provides an effective Hamiltonian for the bath whose ground wave function is a coherent state: 
\begin{equation}
\left | g \right \rangle =e^{\sum_{k}\left( \frac{f \lambda_k}{\omega_q} a_k-h.c.\right)} \left | 0 \right\rangle.
\label{eq:ad2}
\end{equation}
Here $\left | 0 \right\rangle$ is the bosonic vacuum state and $f=\left \langle \varphi \left| \left[ (1-\gamma)\sigma_{+} + \gamma \sigma_{-} \right ] \right| \varphi \right \rangle$. At this stage, the bosonic state $\left | g \right\rangle $ can be used to obtain an effective Hamiltonian, $H_{eff}$, for the spin. It is straightforward to show that $H_{eff}=-\frac{\tilde{\Delta}}{2} \sigma_x + C$, where $\tilde{\Delta}=1+\sum_{k} \frac{2 f \lambda^2_k}{\omega_k}$ and $C=\sum_{k} \frac{f^2 \lambda^2_k}{\omega_k}$, i.e. the adiabatic approximation leads only to an enhancement of the transverse magnetic field along $x$ axis, which is, however, enough to interpret the QMC results at $\gamma=\frac{1}{2}$, shown in Fig. \ref{fig1}. 

\begin{figure}[t]
\begin{center}
\includegraphics[angle=0,scale=0.32]{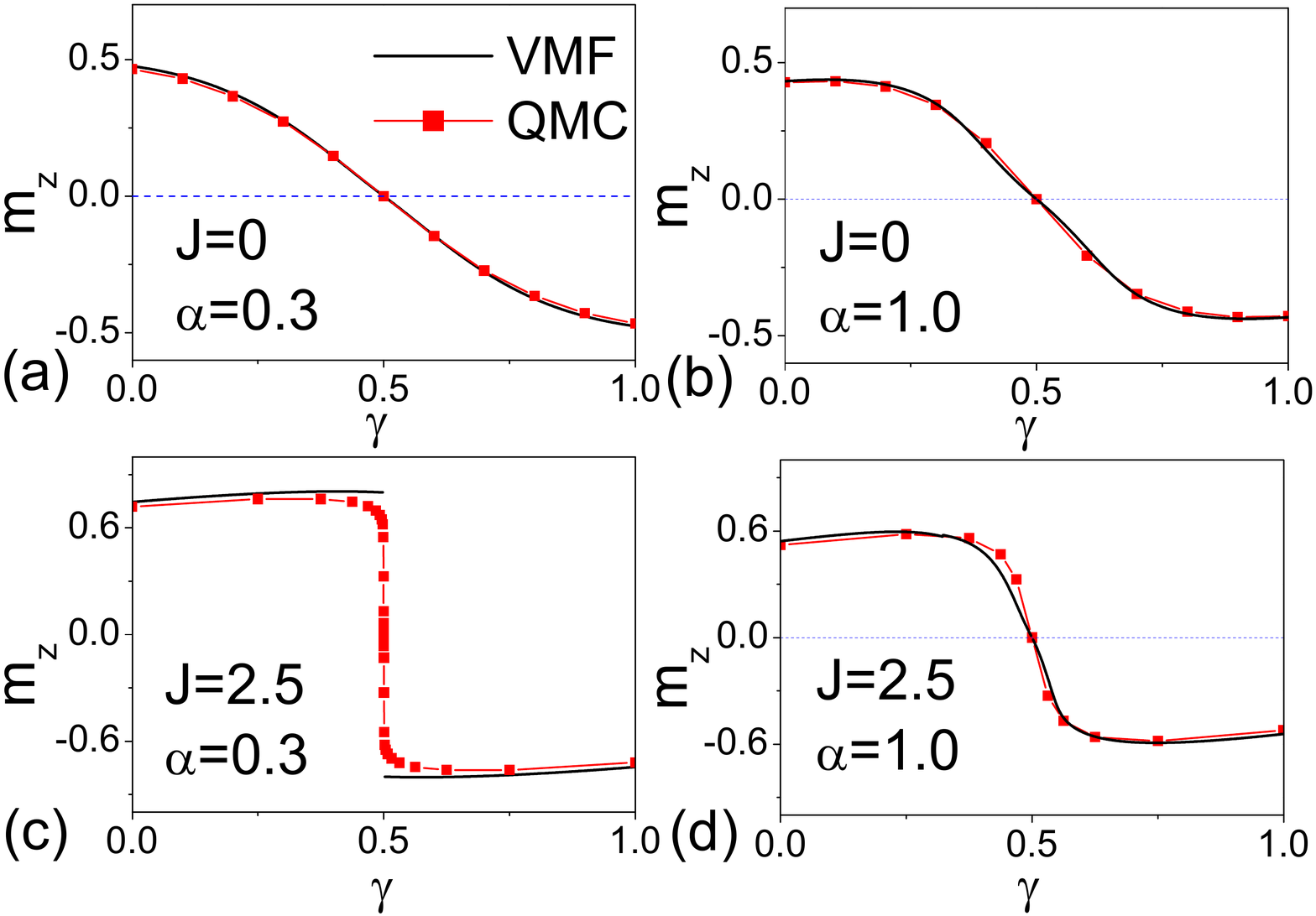}
\caption{Magnetization $m_Z$ as a function of $\gamma$ for different values of $J$ and $\alpha$ by using both QMC and VMF methods.  QMC simulation parameters: $\beta=100$; $L=1$ for (a) and (b),  $L=40$ for (c), and $L=10$ for (d).} \label{fig3}
\end{center}
\end{figure}

In order to get a proper inclusion of the non-adiabatic contributions, relevant for $\gamma \neq\frac{1}{2}$, we first apply an unitary transformation $H_1=e^{S_1} H e^{-S_1}$, with $S_1=-\sum_{k}\left( \frac{f  \lambda_k}{\omega_k} a_k-h.c.\right)$, as suggested by Eq. (\ref{eq:ad2}). The transformed Hamiltonian assumes the form:
\begin{equation}
  H_1= H_{eff}+\sum_{q}\omega_k a^{\dagger}_k a_k + \tilde{H}_I, 
  \label{eq:ad3}
\end{equation}
where $\tilde{H}_I=\sum_{k} \lambda_k A a_k +h.c$, $A$ being an operator acting only on the spin subsystem: $A= \left[ (1-\gamma)\sigma_{+} + \gamma \sigma_{-} \right ] - f$. By treating $\tilde{H}_I$ as perturbation, one includes the non adiabatic contributions. The first order of the perturbation theory, followed by the assumption of no correlation between the emission of different bosons, suggests the following trial state for the bath: $e^{-S_2}\left | 0 \right\rangle$, where $S_2=-\sum_{k} \sigma_z \left( \frac{f_1 \lambda_k}{\omega_k+\tilde{\Delta}} a_k-h.c.\right)$ and $f_1=\gamma-1/2$. In other words, $S_2$ takes into account the possibility that the bath can follow instantaneously the spin oscillations. The effective Hamiltonian for the spin turns out to be:
\begin{equation}
 H_{eff}=-\frac{\Delta_{eff}}{2} \sigma_x +  C_{eff} +h_{eff} \sigma_z ,
\label{eq:ad4}
\end{equation}
where $\Delta_{eff}=\left(\tilde{\Delta}+\sum_{k} \frac{4 f_1^2 \lambda^2_k}{\omega_k+\tilde{\Delta}}\right)e^{-\sum_k \frac{2 f_1^2 \lambda_k^2}{(\omega_k+\tilde{\Delta})^2}}$, $C_{eff}=C+\sum_{k} \frac{\omega_k f_1^2 \lambda^2_k}{\left ( \omega_k+\tilde{\Delta}\right )^2}$, and 
$h_{eff} =2 f \sum_{k} \frac{f_1 \lambda^2_k}{\omega_k+\tilde{\Delta}}$.
It is evident that the quantum fluctuations of the bath induce an effective magnetic field $h_{eff}$ along $z$ axis that breaks the $Z_2$ symmetry. Moreover, the field  $h_{eff}$ changes sign with crossing $\gamma=\frac{1}{2}$. 

The VMF approach can be further improved by by diagonalizing the original Hamiltonian in the subspace spanned by the following $2 M$ wavefunctions: $\psi_i=e^{\sum_{k}\left(l_{k,i} a_k-h.c.\right)}\left | 0 \right\rangle \left | \uparrow \right\rangle$ and $\phi_i=e^{\sum_{k}\left(h_{k,i} a_k-h.c.\right)}\left | 0 \right\rangle \left | \downarrow \right\rangle$, with $i=1, .... M$, in analogy with approaches for the Holstein model \cite{giulio2,giulio3,nocera,perroni}. Here, as suggested by the above described approach, we assume: $l_{k,i}=\frac {f \lambda_k}{\omega_k} + \frac{\lambda_k f_i}{\omega_k+\Delta_i}$, $h_{k,i}=\frac {f \lambda_k}{\omega_k} - \frac{\lambda_k f_i}{\omega_q+\Delta_i}$, being $f$, $f_i$ and $\Delta_i$  $2 M+1$ variational parameters. In Fig. \ref{fig3}(a) and \ref{fig3}(b), where we successfully compare the calculated magnetization $m_Z$ with QMC results for the single spin, we prove that $M=2$ is enough to get a very accurate description of the spin polaron in any regime. Moreover,  in the case of a single spin, as expected, the magnetization vanishes at $\gamma=\frac{1}{2}$, and,  as a function of $\gamma$, shows a crossover behaviour, weakly dependent on  $\alpha$,  from positive to negative values. 

In the case of a spin chain, as shown in Fig. \ref{fig3}(c) and (d), the calculated magnetization $m_Z$ perfectly matches QMC results for $J$ larger than $J_C$ (panel (c), $\alpha=0.3$, $J_C=2.47$ at $\gamma=\frac{1}{2}$) and smaller than  $J_C$ (panel (d), $\alpha=1$, $J_C=3.63$ at $\gamma=\frac{1}{2}$). In the first case, there is an actual first order transition induced by $\gamma$ between states with opposite magnetization, while, in the second case,  only a crossover takes place at $\gamma=\frac{1}{2}$. We remark that the first order transition occurs also for lower values of $\alpha$, therefore the spin system  shows a large sensitivity to external perturbations even at minimal coupling \cite{Suppl2}. 

\begin{figure}[t]
\begin{center}
\includegraphics[angle=0,scale=0.30]{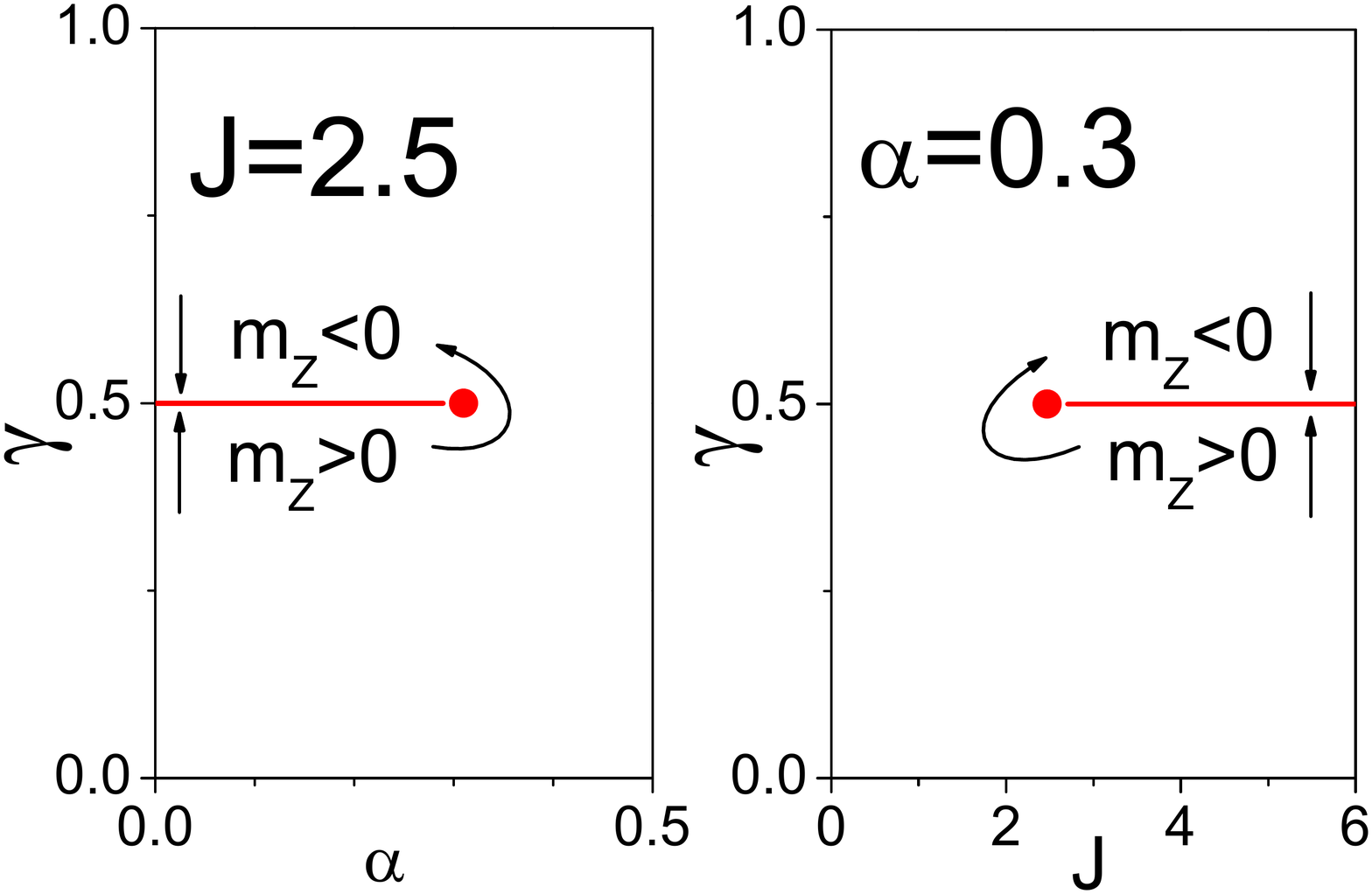}
\caption{Phase diagram with $\gamma$ vs. $\alpha$ at $J=2.5$ (left panel), with $\gamma$ vs. J at $\alpha=0.3$ (right panel). Vertical arrows indicate first order transitions, circles denote second order critical points which can be encircled (curved arrows).  QMC simulation parameters: $\beta=100$, $L=40$.} \label{fig4}
\end{center}
\end{figure}

Finally, we complement the phase diagram of Fig. \ref{fig1}, showing in Fig. \ref{fig4} the diagram with $\gamma$ vs. $\alpha$ at fixed 
$J$ (left panel ), and  the diagram with $\gamma$ vs. $J$ at fixed $\alpha$ (right panel). It is apparent that one gets an horizontal line at 
$\gamma=\frac{1}{2}$ of first order transitions terminating with a second order critical point (denoted with a circle in the panels of  Fig. \ref{fig4}). From the comparison with the phase diagram in Fig. \ref{fig1}, the critical point at $\alpha=0.3$ in the right panel corresponds to $J_c=2.47$.  As expected in this kind of transition \cite{sachdev}, one can go continuously from  a magnetization state to its opposite encircling the second order critical point (curved arrows). 

{\it Conclusions.}
In this letter,  through a numerically exact QMC method and VMF approach, we have shown that  a combination of  Jaynes-Cummings and anti-Jaynes-Cummings spin-bath  couplings drive the transition in a transverse field Ising chain from the second to the first order even for a low dissipation strength. These features are present only for quantum baths. We remark that, in addition to spin-spin interactions \cite{bernien,keesling}, Jaynes-Cummings couplings can be implemented in quantum simulators \cite{pedernales,braumuller}, which, therefore, can potentially test  the results exposed in this work. Moreover,  the spin system is not only sensitive to the properties of non-classical  baths, but also to weak magnetic fields, implying that  the set-up proposed in this work  can be also relevant for the realization of small-scale quantum sensors. Finally, the approaches used in this work can be easily generalized to more complex spin systems \cite{jin2,weber} and spin-bath couplings \cite{zheng}, and in principle allow also to include the effects  of coherent sinusoidal drives \cite{landa}. 

\begin{acknowledgments}
V.C. and G.D.F. acknowledge a critical reading of the manuscript by N. Nagaosa.
\end{acknowledgments}


\begin{thebibliography}{100}
\bibitem{check} We have checked that our method perfectly reproduces the QMC results reported in
\cite{werner1}.
\end{thebibliography}
\end{document}